# GravityCam: Higher Resolution Visible Wide-Field Imaging

Craig Mackay[a], Martin Dominik[b], Iain Steele[c]
[a]Institute of Astronomy, Madingley Road, Cambridge, CB3 0HA, UK;
[b]University of St Andrews, UK; [c]Liverpool John Moores University, UK.

## ABSTRACT

The limits to the angular resolution achievable with conventional ground-based telescopes are unchanged over 70 years. Atmospheric turbulence limits image quality to typically ~1 arcsec in practice. We have developed a new concept of ground-based imaging instrument called GravityCam capable of delivering significantly sharper images from the ground than is normally possible without adaptive optics. The acquisition of visible images at high speed without significant noise penalty has been made possible by advances in optical and near IR imaging technologies. Images are recorded at high speed and then aligned before combination and can yield a 3-5 fold improvement in image resolution. Very wide survey fields are possible with widefield telescope optics. We describe GravityCam and detail its application to accelerate greatly the rate of detection of Earth size planets by gravitational microlensing. GravityCam will also improve substantially the quality of weak shear studies of dark matter distribution in distant clusters of galaxies. The microlensing survey will also provide a vast dataset for asteroseismology studies. In addition, GravityCam promises to generate a unique data set that will help us understand of the population of the Kuiper belt and possibly the Oort cloud.

## 1. INTRODUCTION

Over the years, astronomers have discovered a great deal from using survey data from a variety of sources. One of the first was the original Photographic Sky Surveys carried out by dedicated Schmidt telescopes in both North and South Hemispheres [1]. Today the Sloan Digital Sky Survey (SDSS) [2] made with a 2.5 m telescope with a very wide field of view has provided high quality digital data of a significant part of the northern sky. The Dark Energy Survey (DES - https://www.darkenergysurvey.org/) has been carried out on a modified 4 m class telescope in Chile using a large area detector to image a substantial part of the southern hemisphere. Additional surveys are planned. For example, the Large Synoptic Survey Telescope [3] which is in an early stage of construction, also in Chile, uses an 8 m class telescope with a 3.5° field of view to image large areas of the sky repeatedly with relatively short exposures. One of the principal programs of the LSST will be to survey much of the sky every few nights to allow the detection of exploding supernova in distant galaxies as well as earth-approaching objects.

There have been very significant improvements in the quality of detectors used for astronomy and in particular the charge coupled devices (CCDs). The quality, consistency and repeatability of these devices has allowed large area detector arrays to be built with quantum efficiencies approaching 100%. Detectors are available with broad spectral response and low read-out noise. What has not changed in the last 70 years is the capacity of these surveys to deliver images any sharper than those of photographic plates. Improvements in detector technology have allowed us to detect objects at ever greater distances but increasingly those objects are seen as essentially unresolved by ground-based telescopes. We know from Hubble Space Telescope images that higher angular resolution makes an enormous difference to our capacity to understand these remarkable objects. Even within our own Galaxy there are many regions where confusion dominates because stars are so close together. Advances in adaptive optics technologies have allowed higher resolution images to be obtained over very small fields of view, typically a few arcsec in the visible and therefore of no help in delivering sharper images over a substantial field of view. It is this deficiency that GravityCam is intended to overcome. Our goal is not to produce diffraction limited images but simply images that are sharper than normally obtained from the ground. There are several key scientific programmes that will benefit substantially from such an instrument and are described below in more detail. One of the most exciting is the detection of good numbers of Earth-mass exoplanets by surveying tens of millions of stars in the bulge of our own Galaxy. We can also track the distribution of dark matter in distant clusters of galaxies by looking at the distortions in galaxy images. GravityCam can provide a unique new dataset by surveying millions of stars with high time resolution to enable astroseismologists to better understand the structure of the interior of those stars. It can also allow a detailed survey of Kuiper belt objects.

[a] cdm "at" ast.cam.ac.uk

## 2. SUMMARY OUTLINE OF GRAVITYCAM

One of the most effective ways of improving the detail in astronomical images is to use some of the methods of Lucky Imaging [4] [29]. Here we take images rapidly (10-30 Hz), and identify the tip-tilt offset by measuring the position of a bright compact object in the field. This technique almost completely eliminates the tip tilt distortions caused by atmospheric turbulence. The turbulent power spectrum is strongest on the largest scales. In this way we can improve the median seeing measured on a typical astronomical observatory by a factor between 2.5 and 3. The next level of disturbance of the wavefront entering the telescope is defocus. If we use a smaller fraction of the images and add the sharpest images together we find that image resolution is improved further to a total of a factor of 4 with a 50% selection. Even more restrictive selection can ultimately give even higher factors. However our scientific programmes do not need these more restrictive selection.

With ground-based telescopes of size similar to Hubble (2.5 m) Lucky Imaging can produce close to Hubble resolution (0.1 arcsec) with selections of 20-50%. However as telescopes become larger, the chance of getting a really sharp image become rapidly smaller. It is only possible to get even higher resolution on bigger telescopes using other techniques such as combining Lucky Imaging with low order adaptive optics [5]. These approaches are beyond the scope of this paper. However we propose siting GravityCam on a somewhat bigger telescope such as the NTT 3.6 m in La Silla, Chile (Figure 1). From our experience on that telescope we find that 100% selection yields about 0.3 arcsecond resolution, and 50% selection yields better than ~0.2 arcsec. This site at La Silla has a median seeing of about 0.75 arcsec.

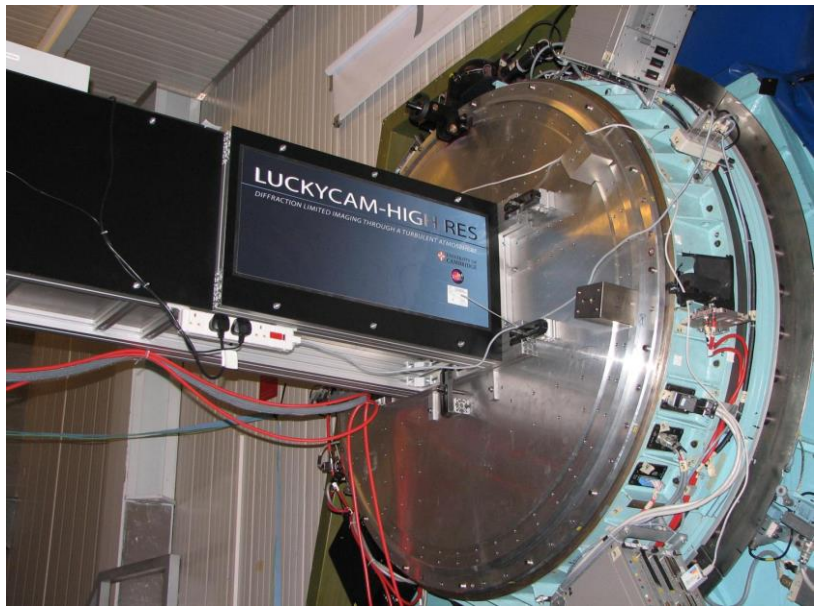

Figure 1: Prototype of the GravityCam detector mounted on one of the Naysmith platforms the NTT 3.6 m telescope of European Southern Observatory in La Silla, Chile. This consisted of a single EMCCD behind a simple crossed prism atmospheric dispersion corrector being run in the standard lucky imaging mode. There is considerable space to mount an instrument on the telescope which has extremely high optical quality and is located in a top astronomical site.

The key requirement is that the detectors used must be able to run at frame rates >10 Hz with negligible readout noise. This will enable very faint targets to be detected. Recently the development of electron multiplying CCDs made this possible and EMCCDs and were quickly taken up for astronomy [6]. When operated conventionally, EMCCDs have a rather high readout noise. However there is a gain mechanism within the detector that allows that to be overcome and allow photon counting operation. Even if photon counting is not needed, it is possible to reduce the read noise to a level that is acceptable. There has been great progress more recently in the development of high-performance CMOS devices. With the right kind of architecture, CMOS devices been be made with very low read-out noise levels (~1 electron RMS) while running at fast frame rates (10-30 Hz). These CMOS devices have many of the excellent characteristics of CCDs such as high quantum efficiency, good cosmetic quality, and high and linear signal capacity. CMOS devices have a further big advantage in that they are capable of being butted together on 2 or 3 edges. EMCCDs have a significant part of the silicon area devoted to a storage region and the form of the device requires a relatively wide area around the device for the fast parallel clocking electrodes. In practice, EMCCDs will only cover about 1/6 of the area with sensitive silicon but this means that to fully cover a field GravityCam has to combine 6 individual pointings. In comparison, with CMOS devices, it is possible to get up to values in excess of 85% and survey an entire field in a single pointing.

## 3. PLANET DEMOGRAPHICS DOWN TO LUNAR MASS ACROSS THE MILKY WAY THROUGH GRAVITATIONAL MICROLENSING

Within foreseeable time, gravitational microlensing [7] [8] remains the only approach suitable to obtain population statistics of cool low-mass planets throughout the Milky Way, orbiting Galactic disk or bulge stars (two populations with notably different metallicity distributions). A GravityCam microlensing survey will provide sensitivity down to Lunar mass, a territory widely considered to be reserved for space missions like WFIRST [9].

The gravitational microlensing effect is characterised by the transient brightening of an observed star due to the gravitational bending of its light by another star that happens to pass in the foreground. This leads to a symmetric achromatic characteristic light curve (see Fig. 2), whose duration is an indicator of the mass of the deflector.

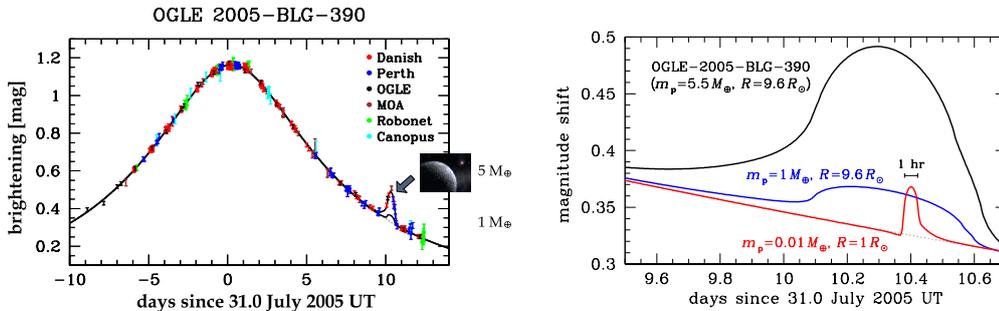

Figure 2: (left) Model light curve and data acquired with 6 different telescopes of microlensing event OGLE-2005-BLG-390, showing the small blip that revealed planet OGLE-2005-BLG-390Lb [31] with about 5 Earth masses. An Earth-mass planet in the same spot would have led to a 3% deviation. (right) Signature of planet OGLE-2005-BLG-390Lb with mp = 5.5 M⊕ and a source star with R = 9.6 R☉, together with those for an Earth-mass planet in the same spot, and a Lunar-mass body with a Sun-like star. Even the latter would be detectable with 2% photometry and 15 min cadence [10]

A planet orbiting the foreground ('lens') star may reveal its presence by causing a further signature that lasts between days for Jupiter-mass planets down to hours for planets of Earth mass or below [11] [12]. Shorter signals do not arise because of the finite angular size of the source star, whose motion relative to the foreground 'lens' star limits the signal amplitude by smearing out the effect that would arise for a point-like source star [13] [14]. Extending the sensitivity to less massive planets therefore means to go for smaller (and thereby fainter) source stars [9]. GravityCam will make the crucial difference by resolving main-sequence stars in the crowded fields of the Galactic bulge (see Fig. 3). We can go about 4 magnitudes deeper than the current state-of-the-art survey OGLE-IV [15] for the same signal-to-noise ratio and exposure time of 2 min, achieving ≤ 5% photometry for the full range 19 < I < 22. With stars at I ~ 16 being about 10 times larger than stars at I ~ 20, we gain a factor 100 in planet mass at the same sensitivity (see Fig 4).

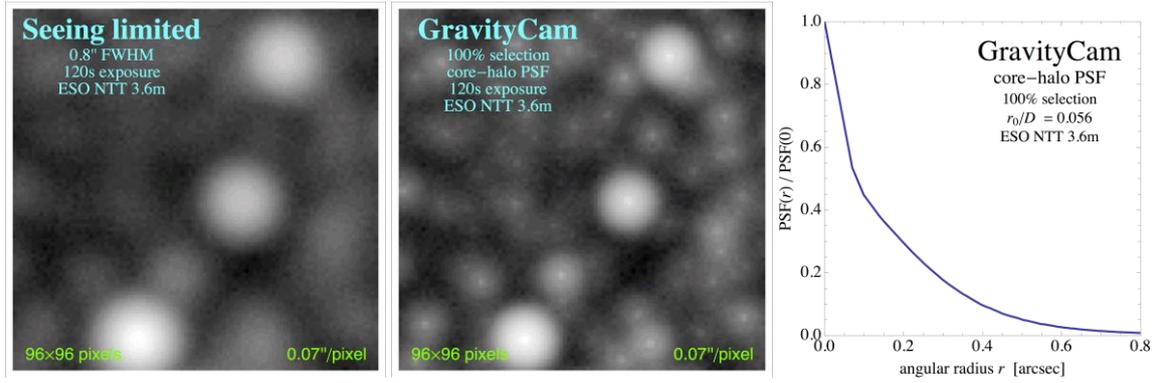

Figure 3: Simulated ESO NTT images of about 7''×7'' size for 2 min exposures, showing the improvement resulting from GravityCam as compared to being limited by an average 0.8'' FWHM, where the core-halo point spread function for 100% frame selection shown on the right has been adopted, while GravityCam gives 0.07''/pixel.

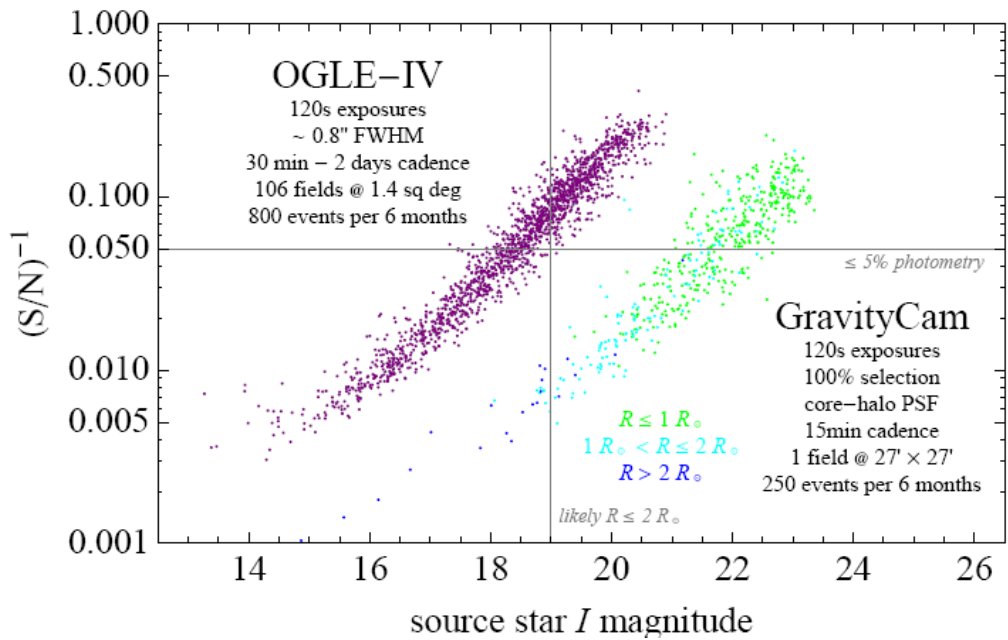

Figure 4: Comparison of performance between OGLE-IV and a microlensing survey with GravityCam on the ESO NTT that uses the same exposure time of 2 min. A single field of 27' × 27' can be monitored at 15 min cadence, sufficient to characterise deviations even by Lunar-mass bodies. While OGLE-IV misses out on providing ≤ 5% photometry on main-sequence source stars, small variations in the brightness of such small stars can be well-monitored with GravityCam.

## 4. TRACING THE DARK MATTER IN THE UNIVERSE BY GRAVITATIONAL LENSING

The fact that matter bends electromagnetic radiation, as predicted by Einstein's General Theory of Relativity [7] and known as 'gravitational lensing', offers a unique opportunity to study the distribution of Dark Matter in the Universe, which is key to our understanding of structure formation. The matter content of galaxy clusters can be inferred from two major signatures: *strong* lensing, providing multiple, highly-magnified and distorted images of distant cosmic sources, and *weak* lensing, giving a sample of single slightly distorted images of many sources for statistical analysis [16] [17].

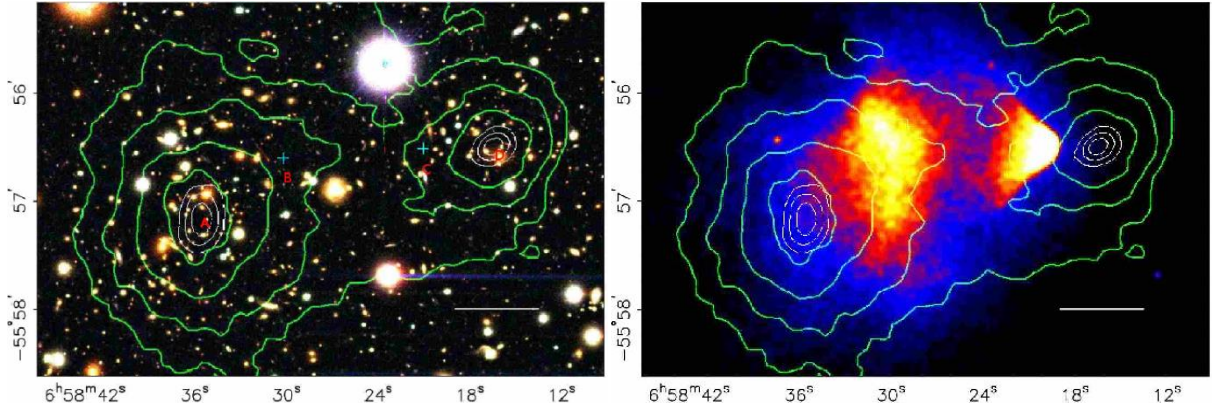

Figure 5: The Bullet Cluster; left panel: optical image with contours showing projected mass derived from lensing; right panel: same lensing mass map contours now with X-ray image showing location of hot gas (dominant component of normal matter). As found by [18], the mass budget is dominated by dark matter. The projected mass derived from lensing with the HST is excellent but with ground based studies it is extremely hard to recover with any accuracy.

Our ability to quantify these signatures crucially depends on the angular resolution of the underlying images, while accuracy of the lens mass reconstruction also depends on the number density of distant background sources. In fact, this technique is calling for the deconvolution of images only slightly bigger than the atmospheric seeing size of ~1-2 arcsec when looking for distortions that are perhaps 1/10 of that size. GravityCam would not only clearly outperform any other ground-based instrument by providing high angular resolution (~0.15'') for faint objects over a wide field, but also compare favourably to space-based facilities. Its resolution and sensitivity would be comparable to the ACS instrument on HST, while covering a wider field of view (see Fig. 5).

A specific instrumental challenge arises from the so-called brighter-fatter-effect (BFE), referring to the phenomenon that charge coupled device detectors produce bigger when the object is brighter [19]. This affects reference stars used for the deconvolution process and also changes the apparent ellipticity of the image significantly. We expect this to be an aspect on which GravityCam will fare better than ESA's Euclid spacecraft [20], a 1.2 m diameter telescope with an array of CCD detectors covering a very similar field of view. With Euclid being located at the L2 position, communication limitations mean that exposure times will be relatively long so the images are likely to suffer to some extent from the BFE.

The other significant difference between GravityCam on the ESO NTT and Euclid is that the NTT has 9 times the collecting area, while the sky in space is not that much darker in the visible than it is from the ground on a good site. GravityCam would enable an unprecedented number of distant galaxies to be used, permitting a more far-reaching three-dimensional description of the dark matter distribution in the Universe.

GravityCam will be ideally suited for studies of cluster substructure, along with observed higher-order shape distortions of galaxy images [21], overcoming current respective limitations of space-based and ground-based facilities and combining their advantages (see Fig. 6).

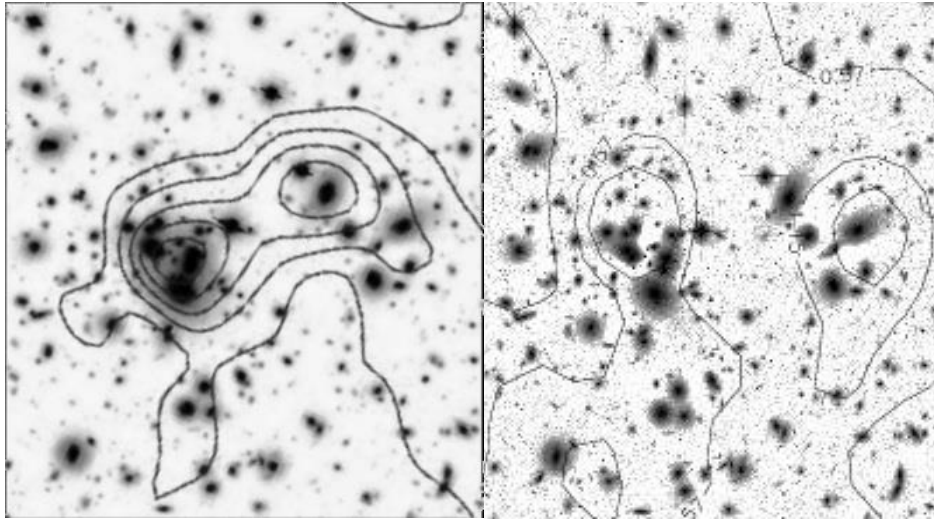

Figure 6: Analysis of the distortion signal to determine the distribution of substructure in the galaxy cluster Abell 1689 using a ground-based wide-field image from Subaru [22] (left) and a deep space-based HST/ACS image (right) [23] with contours showing the reconstructed mass distribution. GravityCam images would be both of high angular resolution and deep (as the space-based data), and at the same time cover a large area like the ground-based data (tens of arc minutes compared to a few arc minutes as HST/ACS).

## 5. GRAVITYCAM: DETAILED DESCRIPTION OF THE INSTRUMENT.

The largest telescopes are generally designed to deliver relatively narrow fields of view. However, for a survey instrument field of view is important. GravityCam may be mounted on any telescope with a wide field of view. Typical detector pixel sizes are well matched to a focal plane scale in the region of 5 arcsec/mm. We suppose here that GravityCam is mounted at the Naysmith focus of the NTT telescope of the European southern Observatory at La Silla Chile. This telescope has a flat 0.5° by 0.5° field of view with Ritchey-Chretien optics. The plate scale is 5.36 arcsec/mm. This means there is no need to use any reimaging optics and the detector is placed essentially in the normal focal plane of the telescope. We use an array of detectors close-packed and operating in synchronism to minimise inter-detector interference (Fig. 7). In order to give good quality images free from residual chromatic aberration from the atmosphere light first goes through an atmospheric dispersion corrector. This is particularly important in crowded field imaging as well as where accurate measurement of the shape of the galaxy is critical. Telescopes such as the LSST [3] that hope to avoid using an ADC are likely to have significant problems particularly with studies such as the weak shear gravitational lensing program described above. In front of the detectors may be mounted interchangeable filter units should they be required. It is worth mentioning that gravitational lensing is a completely achromatic process and, unless there is a good scientific case to use a filter, filters may be dispensed with in order to give the highest sensitivity for the survey in question. However in many cases it will be important to get detailed information of the stars and galaxies being targeted. That will require filters that give that capacity to be available. For example, in order to understand the internal structure of stars, asteroseismologists need to know the colours, effective temperature, log(g) and metallicities of the stars being studied. This will require the availability of appropriate filters for these measurements.

The instrument enclosure of GravityCam is mounted on the NTT via its standard Naysmith platform image rotator (see Figure 1). Otherwise only a filter changer is needed. The detectors are cooled to minimise dark current to a temperature between -80 and -100°C. The detector package will therefore need to be contained within a vacuum dewar enclosure, and cooled with a recirculating chiller or with liquid nitrogen. Driver electronics will be needed for each detector unit and these need to be designed in a modular fashion to allow easy replacement in case of module failure. The volume of the entire GravityCam package mounted on the Naysmith platform of the NTT might be about one cubic metre. The computer system necessary to serve the large number of detectors would be very much bigger, but does not need to be located particularly near to the detector package.

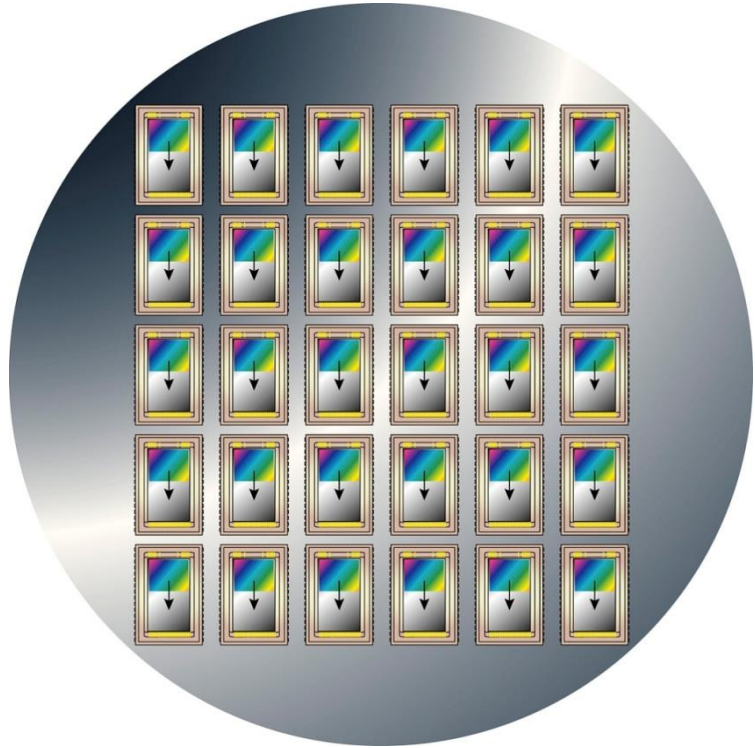

Figure 7: The detectors in GravityCam consist of a close packed array to fill completely the field of view of the telescope. The sensitive parts of the EMCCDs shown diagrammatically here only cover about 1/6 of the area of the detector package. Unfortunately it would be very difficult to re-engineer such devices to give substantially better coverage. Each detector has an image area plus a near-identical storage area. CMOS detectors have the attraction of offering at least 85% fill factor with current manufacturing technologies. CMOS imaging sensors are ubiquitous for many applications such as web cams and mobile phones, but not yet on large telescopes. However CMOS devices under development for astronomy are making rapid progress so are highly likely to be available for use in GravityCam.

## 6. GRAVITYCAM: DETECTOR PACKAGE.

The ability to work at high frame rate with good readout noise is key for the detector package. In order to deliver a significant improvement in angular resolution then a minimum detector frame rate is probably around 10 Hz, and the frame rate closer to 30 Hz is preferable. This permits individual images to be checked for image quality and, if subsets are to be combined separately, the quality selection process carried out. Ideally, higher frame rates enable the system to work under a wider range of observing conditions. However the detector frame rate has considerable effect on the processing requirements of the computer system. There is a further complication when observing fields at high galactic latitude as there are likely to be fields relatively empty of reference stars by which to judge the quality of each frame. The higher frame rates will reduce the signal-to-noise on the reference stars and, at those latitudes, that may make it harder to achieve the necessary performance without reducing the frame rate. However in most fields a simple cross correlation between the reference frame (rather than a singular specific reference object) and the new frame will allow the tip tilt errors to be eliminated. The quality of that cross correlation can also be used to measure the quality of the new frame.

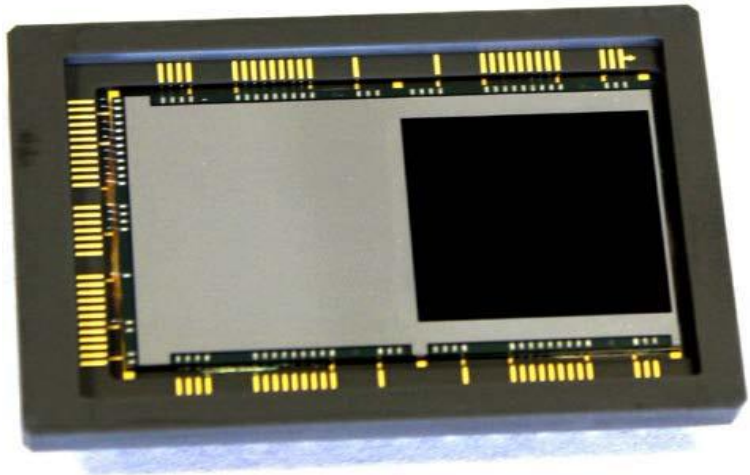

Figure 8: the CCD201 (E2V Technologies, UK) is currently the largest area EMCCD capable of working at frame rates above 10 Hz. It has 1024 x 1024 pixels of 13 µm². The internal gain register allows operation with essentially zero readout noise at the expense of much reduced full well capacity. However at fast frame rates this is much less critical.

The fields targeted for gravitational microlensing are full of stars that make the registration easy to achieve. Nevertheless detector selection requires a careful trade-off. One possible choice for the detector is an electron multiplying CCD (EMCCD: Figure 8). These have the advantage of being able to work with internal gain that can produce images with essentially zero readout noise. The gain necessary to achieve that is typically several hundred and that will reduce the maximum full well capacity that may be used for the detector. This reduces the detector saturation level, although at high frame rates detector saturation is much less serious. The way that the gain mechanism works inside an EMCCD adds noise to the signal which effectively reduces the detective quantum efficiency of the detector by a factor of 2. The essentially zero readout noise of an EMCCD does allow it to work when the background sky brightness is a very low particularly at higher frame rates and when working at shorter wavelengths than I-band. One significant side effect of selecting EMCCDs is the relatively small part of the package area that is sensitive to light. This is because of the basic geometry of the device with its image and store structure together with the package that means only 1/6 of its area is light-sensitive.

Conventional CCD architectures will give a much too high readout noise levels when operated at the frame rates we need. There are now high quantum efficiency CMOS detectors with very low read-out noise, <1 electron rms. In principle they have a number of advantages. The frame transfer architecture of an EMCCD is such that when the integrated image is transferred rapidly into the storage region a bright object in the field produces a faint trail of the image that will complicate the photometry. The architecture of CMOS devices means that they do not suffer from this (Figure 9). They use active pixel sensors which stores the charge and then, on command, transfers that signal into electronic components buried underneath the sensitive silicon. While that charge is being read out the active pixel will integrate light for the following frame. Astronomical CCDs have low read-out noise because they are read out very slowly. CMOS detectors may be made with integrated signal processing electronics within the device itself. This dramatically simplifies the driver electronics and greatly reduces power dissipation in the vacuum enclosure. CMOS technology is what is used in standard computer processor chips and therefore integrating even rather complicated electronics in a small area is relatively straightforward. Using one analogue processing chain with a single analogue to digital converter for each column of the detector, for example, means that each pixel is read out relatively slowly and excellent readout noise may be obtained (Figure 10). Readout noise levels below one electron rms read noise have been obtained routinely in a number of devices. The latest CMOS devices are available with back illuminated (thinned) architectures and can be made with deeply depleted silicon that allows much higher quantum efficiencies in the far red part of the spectrum. All these capabilities add significantly to the sensitivity and efficiency of the detector package.

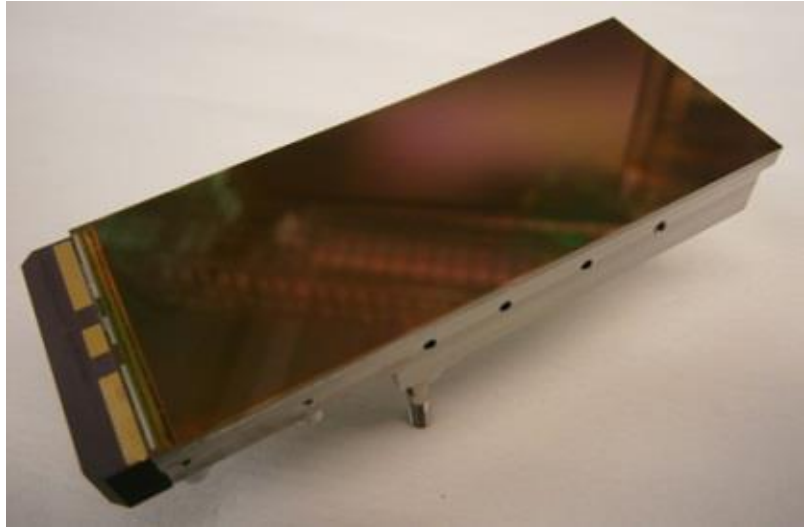

Figure 9: Example of large area CMOS device, the CIS113 from E2V technologies (Chelmsford, UK) [24]. This has 1920 x 4608 pixels, each 16 x 16 µm. It is back illuminated and 3-edge buttable. This device has analogue outputs but other designs are available (Figure 10) with integrated signal processing electronics delivering very low read-out noise. This device is approx. 30 x 80 mm.

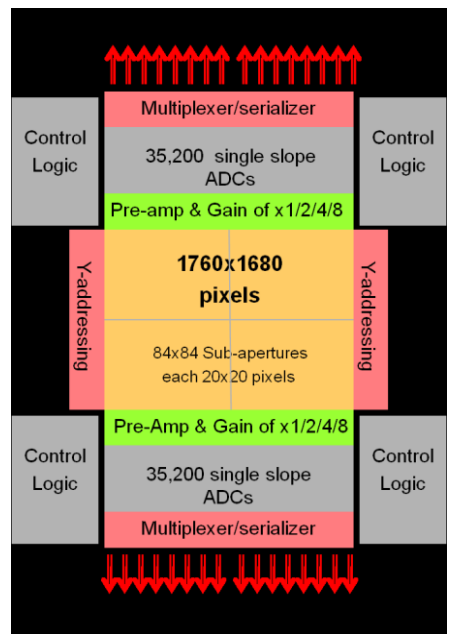

Figure 10: Example of the internal organisation of a highly integrated CMOS sensor for astronomy. The internal architecture is sub-divided into separate areas that may be read out in parallel. This particular device is divided into sub apertures 20 x 20 pixels. These may be read out randomly or sequentially. The signals from the sub apertures are pre-amplified and then passed in this device to 70,400 single slope analogue to digital converters. The digitised data are multiplexed to a total of 88 low-voltage differential serial links to be passed back to the computer [25].

The data from each of the columns of such a device are multiplexed in order to be transmitted back to the processing computer system. Although this all sounds very complicated these processes may be substantially integrated and carried out within the CMOS device leading to a very simple package as far as the engineer employing such devices is concerned. Such a device is driven digitally rather than with the more demanding analogue ones making the implementation of the GravityCam detector package with CMOS devices significantly more straightforward. Finally, it is also important that CMOS devices are now being manufactured so that they may be butted on 2 or 3 out of the 4 sides. This means that the field of view of the telescope is covered much more efficiently with in excess of 85% covering factor. This leads to considerable improvement in efficiency compared to the 16% possible with an EMCCD. The intrinsic structure of the EMCCD makes it very hard to achieve fast buttable detectors without a very expensive custom development programme. The consequence is that an EMCCD based GravityCam covering a full field requires 6 separate pointings. With a CMOS detector then the single pointing will cover 85% of the field which for statistical surveys is all that is necessary. It will also cut down the time used for repeated field acquisitions.

The key performance requirement is good readout noise at the same time as a high frame rate. It is worth noting in passing that there are now near infrared detectors manufactured by Selex [26] that are getting very close to the desired performance in the near infrared. These use mercury cadmium telluride (MCT) sensors and cover the spectral range from about 0.8 µm-2.2 µm [27].

## 7. COMPUTER INTERFACE AND SOFTWARE STRUCTURE.

We have worked with fast EMCCD detectors for astronomy in Cambridge for a number of years and our experience suggests that it will be relatively straightforward to manage at least 10 - 20 Mpixels at 30 Hz frame rate with a single modern data processing computer. Detectors using 13-16 µm pixels and a nearly fully-filled focal plane will produce about 0.5 Gpixels per frame. We have simply assumed that each computer will be a relatively standard well-configured PC type computer. Each computer receives data from the modules over a fibre optic link. Once the relative offset of each image is established it is shifted and added to the summed image. It is usually safer to combine images into different bins depending on the seeing conditions which are highly variable even when the seeing appears to be good. In principle, images may be processed individually but the data volume that produces would quickly become unmanageable.

As the point spread function in each and every frame is very well defined when carrying out the gravitational microlensing survey it is probably a good to strategy to deconvolve each of the images so that PSF halos are further suppressed. This degree of processing requires more significant provision and our models suggest that this is best done with graphics processor unit (GPU) cards. Indeed, if they are to be used it is likely that much of the routine processing is also best done on those cards as the data will be transferred into them anyway. The data processing requirements suggest that we will need an array of 20-50 good-performance PC computers.

In surveys that make repeated visits to the same fields the system will accumulate knowledge of the photometric characteristics of all the objects in each field of view. The GravityCam system will access the archive of detailed existing knowledge of that field to compare with the measurements in real time. Each new image can be compared immediately with the knowledge base that exists already for that field. In the case of the gravitational microlensing survey, the principal data needed for each star is an understanding of its intrinsic variability. When looking for gravitational microlensing events, events typically lasting over days or weeks and follow a well-defined light curve but in a few cases deviations from the standard light curve will be noticed. Such an event must be flagged immediately to ensure that the field is revisited frequently enough to provide the photometric monitoring as necessary. This also enables other instruments or telescopes programs to be alerted to the new event as appropriate. Indeed it would also be possible to change the observing program of GravityCam to provide follow-up photometry.

The operation of GravityCam will be overseen by a supervisory computer system that is responsible for setting up each of the detector modules and checking that all of the technical variables are in place. The supervisory computer will also monitor instrument enclosure temperatures once the devices are running. At that point data taking may be started. For each new field accessed, where appropriate, archival data on that field is loaded into each data processing computer. The supervisory system triggers each data processing computer to get on with the observations of each field. Each image is calibrated photometrically simply by measuring the light from as many reference stars as available to give a photometric calibration of each individual frame. Stars are then tracked and their brightness compared with the existing knowledge of their characteristics. Individual images are then offset laterally to bring them into positional registration. The new images may then be combined with images already summed , and processed to improve the detailed knowledge of the characteristics of each and every star in the field, in turn allowing the archive data to be improved and updated. The exact frequency with which this is done depends very much on a detailed trade-off involving the processing power required and the information needed and of the particular observing program.

## 8. GRAVITYCAM: SENSITIVITY ESTIMATES AND PREDICTIONS.

There is often concern about the photometric accuracy of Lucky Imaging or indeed any approach that selects a subset of images. However each frame is used in its entirety or not depending on the percentage selection criteria. Each frame is as photometrically reliable as any other and the overall sensitivity is simply the same as one would achieve with the long exposure time but reduced by the selection efficiency. As a consequence, with 100% image selection followed by shifting and adding then there is no loss in efficiency. Neither of these detectors (EMCCD or CMOS) has essentially any equivalent of "shutter closed time". In practice it may be that individual images are accumulated into more specific image quality bins such as 5% or 10% windows. These can always be combined later to give the full sensitivity. However if only 50% of the images are to be used then the sensitivity achieved in the final image will be reduced obviously by a factor of 2.

It has already been mentioned that gravitational lensing, both microlensing and weak shear lensing, is an achromatic process. This means that in order to achieve the very highest sensitivity then filters may be dispensed with or perhaps restricted to a long pass filter. It is very unfashionable for astronomers to observe without filters but it has a significant influence on the overall sensitivity. With typical CCD or CMOS sensitivity, dispensing with a filter will improve the signal-to-noise on a faint object by about one magnitude when compared with the I band sensitivity. It is interesting to note that deep Galaxy imaging in 1986 [28] without any filter detected high galaxy surface densities at high galactic latitudes. Higher densities were only measured nearly 20 years later by the Hubble Space Telescope Deep Field and by using very long exposures indeed. It is interesting to note that the Euclid space mission [30] scheduled for launch in 2020 includes an imaging instrument (VIS) with only a very broad band (500-900 nm) filter for just this purpose. As a small diameter telescope it is important that it has as higher throughput as possible.

## 9. GRAVITYCAM: PROJECT COSTINGS ON TIMESCALES.

The cost and effort in building GravityCam will be dominated by detector related work. If an EMCCD approach is used then existing detectors can be purchased relatively quickly. If a CMOS approach is to be taken then there is likely to be a development period to optimise the performance of the detectors specified. With CMOS detectors the electronic integration will be much easier and that in turn will make the design of cooling system easier. With EMCCD detectors a significant drive power is required to clock the charge across the device. Power dissipation within the detector subassembly must be managed with care if it is not to become excessive, and thereby increase the cooling power requirement within the dewar.

The turnkey software package will be a critical piece of design work as it must provide timely and consistent image processing. It is essential that it can reliably alert other telescopes and instruments to start to follow new events. The supervisory software must be structured in a relatively configurable and modifiable way so that it is easy to implement different kinds of strategies for observing programs. It is always important to bear in mind that developing software that responds quickly without causing too many false positives with a widefield imaging survey is breaking new ground and creating the software is likely to be at least challenging.

The EMCCDs that we would use already in production so we know that the cost of covering a 0.5° diameter focal plane will be in the region of €1 million. It is much less easy to quantify the costs of a CMOS detector development programme because these devices are in an earlier stage of development. Also, if we use CMOS detectors, we will be essentially purchasing 6 times as many pixels and the cost of silicon detectors that are thinned will be correspondingly higher. At this stage the cost of these detectors will clearly drive the total cost and so accurate estimates cannot really be made. An approach to start with EMCCDs and upgrade the CMOS devices later date would be possible although the aggregate cost would therefore be significantly greater. We estimate that the project cost using EMCCDs would probably be in the region of €5 million and with CMOS detectors closer to €10 million. Most of the work required is relatively straightforward and could be done within 2.5-3 years provided the detectors are available on the correct timescale. This makes the GravityCam project relatively inexpensive and relatively quick to implement on the telescope.

It has the potential to revolutionise several independent branches of astronomy and provides a unique capability not available on any other telescope, ground or space based or indeed planned in the foreseeable future.

## 10. GRAVITYCAM: CONCLUSIONS.

Developments in imaging detector technology mean that techniques of improving the resolution of images taken on ground-based telescopes are now within reach. All the principles have been demonstrated already and there are many astronomical programs which are now being substantially constrained because of the quality of images that can be delivered even on the best ground-based sites. GravityCam provides a new approach to how this may be done and we have outlined several exciting scientific programmes which would benefit greatly if GravityCam was available.

## ACKNOWLEDGEMENTS.

We have benefited from discussions with a number of colleagues and in particular would like to thank Colin Snodgrass, Michael Hirsch, Uffe Gråe Jørgensen, Markus Hundertmark , Rafael Robolo, Keith Horne, Sarah Bridle, Bruno Sicardy, Daniel Bramich and Khalid Alsubai.

## REFERENCES.

[1] Reid, N. & Djorgovski, S.,1993, ISBN: 0-937707-62-7
[2] Gunn et al, 2006, AJ....131.2332
[3] LSST  "The Large Synoptic Survey Telescope ". Phys.org. May 29, 2015.
[4] Mackay, C. D. et al, SPIE vol 5492,128 [2004]
[5] Law, N M, C.D. Mackay et al, ApJ **692**,924-930 (2009).
[6] Mackay, C D, et al., SPIE Vol 4603A, San Jose, 289 [2001]
[7] Einstein, A., *Sitzungsber. Kgl. Preuss. Akad. Wiss.,* 831 (1915)
[8] Paczyński, B., ApJ 304, 1 (1986)
[9] Bennett, D.P., & Rhie, S.H., ApJ 574, 985 (2002)
[10] Dominik, M. et al., MNRAS 380, 792 (2007)
[11] Mao, S., & Paczyński, B., ApJ 374, L31 (1991)
[12] Gould, A., & Loeb, A., ApJ 396, 104 (1992)
[13] Bennett, D.P., & Rhie, S.H., ApJ 472, 660 (1996)
[14] Dominik, M., Phil. Trans. R. Soc. A 368, 3535 (2010)
[15] OGLE-IV (http://ogle.astrouw.edu.pl)
[16] Treu, T., ARAA 48, 87 (2010)
[17] Hoekstra, H., ARNPS 58, 99 (2008)
[18] Clowe, M et al, ApJ648,109 (2006)
[19] Downing et al, SPIE, 6276,9 (2006).
[20] Laureijs, R. et al., "Euclid Definition Study Report", report ESA/SRE(2011)12 (2011)
[21] Goldberg, D.M., & Bacon, D.J., ApJ 619, 741 (2005)
[22] Okura, Y. et al., ApJ 660, 995 (2007)
[23] Leonard, A. et al., ApJ 666, 51 (2007)
[24] Jorden et al, SPIE.9154E..0MJ (2014).
[25] Downing et al ,SPIE.  Volume 9154.(2014).
[26] Finger et al, SPIE 9909-40 (2016).
[27] Hall et al, SPIE, 9915 (2016)
[28] Hall, P., and Mackay, C.D. MNRAS, 210, 979 (1986).
[29] Baldwin, J.E, Tubbs, R.N., Mackay, C.D., et al, (2001), Ast & Astrophys, vol. 368, L1-4.
[30] Euclid space mission: http://www.euclid-ec.org/
[31] Beaulieu, J-P et al, Nature, 439,437 (2006).